\documentclass[twocolumn,aps,prb,groupedaddress,showpacs]{revtex4-1}
\usepackage{amsfonts}
\usepackage{amsmath}
\usepackage{amssymb}
\usepackage{txfonts}
\usepackage{pxfonts}
\usepackage{graphicx,bm,units,yfonts}
\usepackage[table]{xcolor}

\begin{document}

\title{AIII and BDI Topological Systems at Strong Disorder}
\affiliation{Department of Physics, University of Illinois-Urbana Champaign, Urbana, IL 61801, USA}
\author{Juntao Song$^{1,2}$ and Emil Prodan$^{1}$}
\affiliation{\ $^{1}$Department of Physics, Yeshiva University, New York, NY 10016, USA\\
$^2$Department of Physics, Hebei Normal University, Hebei 050024, China}

\begin{abstract}
Using an explicit 1-dimensional model, we provide direct evidence that the one-dimensional topological phases from the AIII and BDI symmetry classes follow a $\mathbb Z$-classification, even in the strong disorder regime when the Fermi level is embedded in a dense localized spectrum. The main tool for our analysis is the winding number $\nu$, in the non-commutative formulation introduced in I. Mondragon-Shem, J. Song, T. L. Hughes, and E. Prodan, arXiv:1311.5233. For both classes, by varying the parameters of the model and/or the disorder strength, a cascade of sharp topological transitions $\nu=0 \rightarrow \nu=1 \rightarrow \nu=2$ is generated, in the regime where the insulating gap is completely filled with the localized spectrum. We demonstrate that each topological transition is accompanied by an Anderson localization-delocalization transition. Furthermore, to explicitly rule out a $\mathbb Z_2$ classification, a topological transition between $\nu=0$ and $\nu=2$ is generated. These two phases are also found to be separated by an Anderson localization-delocalization transition, hence proving their distinct identity.

\end{abstract}


\maketitle

\section{Introduction}
The topological phases from the AIII and BDI symmetry classes are classified by the winding number $\nu$ of the ground state. The classification table of topological insulators and superconductors \cite{SchnyderPRB2008qy,KitaevArXiv2009oh,RyuNJP2010tq} predicts that non-trivial topological phases exist in the AIII symmetry class for each odd space-dimension. Furthermore, in each odd space-dimension, it is predicted that there are as many topological phases as integer numbers in the additive $\mathbb Z$ group. For the BDI symmetry class, the table predicts a very different scenario. Here, topological phases appear in 0 and 1 space-dimensions, after which there is a void and the next space-dimension where non-trivial topological phases exist is 5. In some dimensions, such as 0, 7 and 8, the classification is $\mathbb Z_2$, while for other dimensions, such as 1 and 11, the classification is $\mathbb Z$. In dimension 5, the classification is $2\mathbb Z$. It has been argued,\cite{SchnyderPRB2008qy,KitaevArXiv2009oh,RyuNJP2010tq} based on the nonlinear-sigma-model and $K$-theory, that this classification remains valid in the regime of strong disorder. However, it is extremely difficult to analyze this regime theoretically because the quantum states cross the Fermi level both from above and below upon deformations of the models, making any argument based on a tiny insulating gap invalid. As always, it is desirable to have a direct confirmation of these predictions, and this is one of our goals for the present work. We focus exclusively on the one-dimensional phases from the AIII and BDI symmetry classes, both predicted to obey a $\mathbb Z$-classification.

An explicit confirmation of the $\mathbb Z$ classification at strong disorder is especially important for the following reason. As shown in Ref. [\onlinecite{MondragonShemArxiv2013ew}], there is an important relation between the winding number $\nu$ and the electric polarization $P$,~\cite{KingSmithPRB1993ho,RestaRMP1994xd,OrtizPRB1994pl,QiPRB2008ng,Hughes2010gh,TurnerPRB2012cu} whenever a chiral symmetry is present (which is the case for both AIII and BDI classes). The later is defined only by modulo integers, and due to the chiral symmetry, it can only take the values 0 and $\frac{1}{2}$. The relation between the two is:
\begin{equation}
P=\nicefrac{1}{2}\ (\nu \ \mbox{mod}\ 2).
\end{equation}
This brings up a set of interesting and important questions. Since the electric polarization is one of the possible physical manifestations of the topological character, and this physical observable takes only two stable values, then: 1) Does $\nu$ itself take only two stable values (0 and 1)? 2) If $\nu$ takes more than two stable values, is it true that all the corresponding topological phases are macroscopically distinct? For example, if $\nu$ is only relevant to modulo 2, then the phase $\nu=2$ would be trivial and we could cross between $\nu=2$ to $\nu = 0$ phases  without going throughout a quantum transition.

To conclusively demonstrate the $\mathbb Z$-classification, and entirely eliminate the possibility of a $\mathbb Z_2$-classification, we generate strong disorder models that display topological phases with $\nu = 0$, 1 and 2, and we study the cascade of topological transitions: $\nu=2 \rightarrow \nu = 1 \rightarrow \nu=0$. A sharp change of the quantized $\nu$ is observed at the phase boundaries, accompanied by an Anderson localization-delocalization transition. This shows that indeed the phases $\nu=0$ and $\nu=2$ are topologically distinct from the phase with $\nu=1$. If the classification was by $\nu\mod 2$, then the phases $\nu=0$ and $\nu=2$ will be indistinguishable. To show that this is not the case, we searched in the parameter space for a situation in which the phase $\nu=0$ is in direct contact with the phase $\nu=2$. After turning on the strong disorder, we similarly find an Anderson localization-delocalization phase transition along the entire phase boundary, proving that $\nu=0$ and $\nu=2$ are indeed distinct phases. If one wonders whether the situation could be different for other models, the answer is no. The classification for these classes is by $\mathbb Z$, which implies that there is a unique generator for the entire sequence of phases and, consequently, all models with a given $\nu$ can be deformed into $\nu$-number of copies of this generating model, hence they are equivalent.

Another important issue is the characterization of these topological phases at strong disorder. It has been a widespread belief among our community that the topological invariants are carried by bulk extended states that resist the Anderson localization, and which reside at energies below and above the Fermi level. This characteristic has been indeed observed numerically for several symmetry classes.\cite{Onoda:2007xo,ProdanPRL2010ew,ProdanJPhysA2011xk,XuPRB2012vu,GilbertPRB2012bv,LeungPRB2012vb} However, a recent study\cite{MondragonShemArxiv2013ew} on the AIII symmetry class revealed that, in this particular case, the entire energy spectrum becomes localized as soon as the disorder is turned on, and it stays localized until an Anderson localization-delocalization transition builds up (from this entirely localized spectrum!) while crossing from one topological phase to another. One implication is that the winding number is {\it not carried} by extended states, and instead each localized state carries a small part of it. Another implication is that the divergence of the localization length does not have to happen through the levitation and annihilation of some extended states above and below the Fermi energy (as it always happens in classes A and AII,\cite{Onoda:2007xo,ProdanPRL2010ew,HalperinPRB1982er}) and instead the divergence can develop from an entirely localized spectrum. Ref. [\onlinecite{MondragonShemArxiv2013ew}] made the prediction that this un-characteristic behavior is present in all symmetry classes where  the $E$ is fixed by the symmetry to a given value. The present work demonstrates that this is indeed true for the BDI class of topological insulators.

The rest of this paper is organized as follows: In Sec. \ref{NCWindingNumber}, we introduce the winding number for a chiral symmetry model in a clean limit, report a real space formula of a non-commutative winding number for a homogeneous disordered model, and explicitly present the numerical algorithm used to compute the non-commutative winding number. Further, two AIII and BDI symmetry lattice models are introduced, and some preliminary properties of these two models are discussed in Sec. \ref{LatticeModels}. In Sec. \ref{NCWNStrongdisorder}, we show the numerical results of the non-commutative winding number and localization length for both AIII and BDI symmetry models, and we explicitly discuss the critical phenomena of phase transitions. Based on the numerical computation of the non-commutative winding number and the localization length, in Sec. \ref{PhaseDiagram} the topological phase diagrams for both AIII and BDI symmetry models are given in the presence of disorder. Finally, a conclusion is given in Sec. \ref{Conclusions}.

\section{The Non-Commutative Winding Number: Definition, Characterization and Computation}\label{NCWindingNumber}

The main tool in our analysis is the recently introduced non-commutative winding number\cite{MondragonShemArxiv2013ew} $\nu$, obtained by a real-space reformulation of the $k$-space expression, followed by a proper generalization to the disorder case. This invariant has been already demonstrated to remain quantized and non-fluctuating (from one disorder configuration to another), though only in the $\nu = 1$ topological phase in the AIII symmetry class.\cite{MondragonShemArxiv2013ew} The non-commutative formula of the winding number has the following attributes:

(i) It is well defined in the regime of strong disorder; a operations involved in the calculation make straightforward sense, as opposed to, for example, the twisted boundary conditions method in which derivatives against the twist angles are invoked, but the multitude of states crossing the Fermi level destroy the required differentiability.

(ii) It is self-averaging in the sense that its actual value can be determined from a single disorder configuration, provided the size of the simulation box is large enough.

(iii) It can be evaluated numerically with extreme accuracy and efficiency.\cite{ProdanAMRX2013bn} In the present simulations, the quantization of the non-commutative winding number is exact up to at least six digits of precision.

(iv) It can be rigorously characterized using Non-Commutative Geometry and shown to stay quantized and non-fluctuating as long as the Fermi level, which is fixed at $E=0$, resides in a region of a localized spectrum.\cite{ProdanOddChernArxiv2014}

In the following we review the theory of the winding number in arbitrary odd space-dimension $D$, and present in more detail the numerical algorithm used to compute the non-commutative winding number.

\subsection{Periodic Lattice Models with Chiral-Symmetry}

The generic translational-invariant, multi-band lattice Hamiltonians take the form:
\begin{equation}\label{LatticeH0}
\big ( H_0{\bm \psi} \big)_{\bm x} =\sum_{{\bm y}\in \mathbb Z^D} \hat{t}_{{\bm x}-{\bm y}}{\bm \psi}_{\bm y},
\end{equation}
where $\bm \psi$ is a $N$-component spinor and ${\hat t_{\bm a}}$'s are $N \times N$ matrices with complex entries, such that $\hat t_{-\bm a}=(\hat t_{\bm a})^\dagger$. Here, $N$ represents the number of molecular orbitals per repeating cells. The Hamiltonian has chiral-symmetry if there exists a unitary $N \times N$-matrix $S$ with the properties:
\begin{equation}
S ^2=1, \ S^\dagger = S, \ \mbox{and} \ S H_0 S^{-1}=-H_0.
\end{equation}
An immediate consequence of this symmetry is that the energy spectrum of the Hamiltonian is symmetric to the reflection relative to $E = 0$. The Fermi level is always fixed at $E = 0$ for systems with chiral-symmetry and since we are interested in insulators, we assume the existence of a spectral gap in the energy spectrum of $H_0$ at $E = 0$. This is possible only if $N$ is even.

When translational symmetry is present, the model can be represented in the ${\bm k}$-space, where the dynamics is generated by the family of Bloch-Hamiltonians:
\begin{equation}
H(\bm k):\mathbb C^N \rightarrow \mathbb C^N, \ \ H ({\bm k}) = \sum_{\bm a} \hat t_{\bm a} e^{i {\bm a}\cdot {\bm k}}.
\end{equation}
In this case, the chiral symmetry reads:
\begin{equation}\label{KSymmetryCond}
S H({\bm k}) S^{-1} =- H({\bm k}) .
\end{equation}
One direct consequence of this symmetry is that the Hamiltonian takes the block-form:\cite{Schnyder:2009qa}
\begin{equation}
H(\bm k)=\left (
\begin{array}{cc}
0 & h(\bm k) \\
h(\bm k)^\dagger & 0
\end{array}
\right )
\end{equation}
if one works in the basis provided by the eigenvectors of $S$. Since the matrix $h$ is invertible, one can define a winding number:
\begin{equation}\label{WindingK1}
\nu  = \frac{-(\frac{D-1}{2})!}{(2 \pi i)^{\frac{D+1}{2}} D!} \int_{\mathbb{T}^D}  \mathrm{Tr} \left \{\big (h({\bm k})^ {-1} {\bf d} h ({\bm k})  \big )^D\right \}\in \mathbb Z,
\end{equation}
over the Brillouin torus $\mathbb T^D$. Furthermore, one can use the homotopically equivalent flat-band version of the Hamiltonian,
\begin{equation}\label{TheQ}
\frac{H(\bm k)}{|H(\bm k)|} = \left (
\begin{array}{cc}
0 & q(\bm k) \\
q(\bm k)^\dagger & 0
\end{array}
\right ),
\end{equation}
and define the winding number as:
\begin{equation}\label{WindingK2}
\nu = \frac{-(\frac{D-1}{2})!}{(2 \pi i )^{\frac{D+1}{2}} D!} \int_{\mathbb{T}^D}  \mathrm{Tr} \left \{\big (q({\bm k})^{-1} {\bf d} q ({\bm k})  \big )^D\right \} \in \mathbb Z.
\end{equation}
The advantage of the latter is that $q$ is a unitary matrix, but many times Eq. (\ref{WindingK1}) is easier to compute analytically.

The topological invariant can be formulated directly in the real-space representation.\cite{MondragonShemArxiv2013ew} Indeed, the flat-band Hamiltonian, in the basis provided by the eigenvectors of $S$ (this time viewed as an operator in real-space), takes the following block-form:
\begin{equation}\label{EqQ0}
\frac{H_0}{|H_0|}=\left (
\begin{array}{cc}
0 & Q_0 \\
Q_0^\dagger & 0
\end{array}
\right ).
\end{equation}
The Bloch-Floquet representation of Eq. (\ref{EqQ0}) is nothing else but Eq. (\ref{TheQ}). Then the covariant real-space representation\cite{MondragonShemArxiv2013ew} of the winding number can be obtained by applying the Bloch-Floquet transformation in reverse on Eq. (~\ref{WindingK2}):
\begin{equation}\label{WindingR1}
\nu = \frac{(i \pi)^\frac{D-1}{2}}{D!!} \sum_{\rho \in S_D} (-1)^\rho \ \mathcal T  \left \{\prod_{j=1}^D Q_0^{-1} [X_{\rho_j},Q_0]  \right \},
\end{equation}
where $\mathcal T$ represents the trace per volume, $\bm X = (X_1, \ldots, X_D)$ denotes the position operator, $(X_j \bm \psi)(\bm x) = x_j \bm \psi(\bm x)$, and the summation is over all permutations of the indices. Since Eq. (\ref{WindingR1}) is completely equivalent with Eqs. (\ref{WindingK1}) and (\ref{WindingK2}), the $\nu$ defined in Eq. (\ref{WindingR1}) takes integer values, if evaluated in the idealistic limit of infinite volumes.

\subsection{Homogeneous Disordered Lattice Models with Chiral-Symmetry}

The generic disordered lattice Hamiltonians take the form:
\begin{equation}\label{LatticeHamOmega}
(H_\omega {\bm \psi})_{\bm x}=\sum_{{\bm y}\in \mathbb Z^d} \ \hat t_{{\bm x},{\bm y}}(\omega) {\bm \psi}_{\bm y},
\end{equation}
where $\hat t_{\bm x,\bm y}(\omega)$ are $N \times N$ matrices with complex entries that depend on $\omega$, a random variable from a probability space $\big (\Omega, d \bm P (\omega) \big )$ representing the disorder configurations. The system is said to be homogeneous if:
\begin{equation}
\hat{t}_{{\bm x}-{\bm a},{\bm y}-{\bm a}} (\omega)= \hat{t}_{{\bm x},{\bm y}}(\mathfrak{t}_{\bm a} \omega),
\end{equation}
where $\mathfrak{t}_{\bm a}$'s are probability-preserving, ergodic automorphisms on $\Omega$, implementing the lattice-translations group. This is an important condition, because then the collection $\{H_\omega\}_{\omega \in \Omega}$ defines a covariant family of observables:
\begin{equation}
T_{\bm a}H_\omega T_{\bm a}^{-1}=H_{\mathfrak{t}_{\bm a}\omega},
\end{equation}
for any lattice translation $T_{\bm a}$. The key point is that, if $F_\omega, G_\omega, \ldots$ are covariant observables, then, according to Birkhoff's ergodic theorem, the trace per volume is independent of the disorder configuration (with probability one) and is equal to an average over disorder:
\begin{equation}\label{Birkhoff}
\mathcal T\{F_\omega G_\omega \ldots\} = \int_\Omega d\omega \ \mathrm{tr}_{\bm 0} \{F_\omega G_\omega \ldots\}
\end{equation}
where $\mathrm{tr}_{\bm 0}$ denotes the trace over the molecular orbitals in the first unit cell. Note that Eq. (\ref{Birkhoff}) states that any correlation function involving covariant observables is self-averaging. This is the principle behind the self-averaging of the non-commutative winding number.

A homogeneous disordered system is chiral symmetric if:
\begin{equation}
S H_\omega S^{-1} = - H_\omega \ \mbox{for all} \ \omega \in \Omega.
\end{equation}
Considering again the flat-band Hamiltonian and working in the basis provided by the eigenvectors of $S$:
\begin{equation}
\frac{ H_\omega}{|H_\omega|}= \left (
\begin{array}{cc}
0 & Q_\omega \\
Q_\omega^\dagger & 0
\end{array}
\right )
\end{equation}
with $Q_\omega$ a unitary operator, the natural generalization of the winding number is then:
\begin{equation}\label{WindingR2}
\boxed{
\nu = \frac{(i \pi)^\frac{D-1}{2}}{D!!} \sum_{\rho \in S_D} (-1)^\rho \ \mathcal T  \left \{\prod_{j=1}^D Q_\omega^{-1} [X_{\rho_j},Q_\omega]  \right \}.
}
\end{equation}
The righthand side can be easily seen to take finite values when the kernel of $Q_\omega$ decays exponentially fast, as is the case (on average) when the Fermi level resides in a region of localized spectrum. Based on our previous arguments, the non-commutative formula evidently has the self-averaging property. However, there is no a prior reason to think that the non-commutative winding number defined in Eq. (\ref{WindingR2}) is integer when the spectral gap is closed. Nevertheless, if the Fermi level resides in a region of dynamical localization, this holds true due to the following index theorem proved in Ref.[\onlinecite{ProdanOddChernArxiv2014}]: \smallskip

\noindent {\bf Theorem:} Let $\sum_{j=1}^D X_j \sigma_j$ be the Dirac operator, where $\sigma_i$'s are the $D$-generators of the odd Clifford algebra. Let $\Pi$ denote the projector onto the positive spectrum of the Dirac operator and assume:
\begin{equation}\label{cond}
\int_\Omega d \omega \ |\langle \bm x|Q_\omega|\bm y\rangle| \leq Ae^{-\gamma |\bm x - \bm y|},
\end{equation}
for some strictly positive $A$ and $\gamma$. Then, with probability 1 in $\omega$, $\Pi Q_\omega \Pi$ is a Fredholm operator and
\begin{equation}
\nu = \mathrm{Index} \ \Pi Q_\omega \Pi.
\end{equation}
Furthermore, the Fredholm index on the right is independent of $\omega$ and is invariant against any continuous deformations of the Hamiltonian as long as Eq. (\ref{cond}) is satisfied.\smallskip

Of course, Eq. (\ref{cond}) holds true if the Fermi level resides in the region of an Anderson localized energy spectrum.\cite{ProdanOddChernArxiv2014} A direct prediction of the above result is the fact that $\nu$ can change only if the localization length diverges at $E=0$. This was already confirmed numerically in Ref. [\onlinecite{MondragonShemArxiv2013ew}] for the crossing $\nu=0 \rightarrow \nu=1$ in a model from AIII-symmetry class, and it will be further confirmed by our numerical results in a broader setting.

\subsection{Computing the Non-Commutative Winding Number}

We now come to the important question of how to compute the non-commutative winding number. For this type of calculations, the following practical solution was devised in a series of works.\cite{ProdanPRL2010ew,ProdanJPhysA2011xk, ProdanAMRX2013bn} The most difficult part of the problem is how to represent or approximate the commutators $[X_j,Q_\omega]$ on a finite volume with periodic boundary conditions. On the infinite volume:
\begin{equation}\label{Comm}
\langle x_j | [X_j,Q_\omega]|x'_j \rangle = (x_j -x'_j) \langle x_j | Q_\omega|x'_j \rangle,
\end{equation}
and clearly the factor $x_j -x'_j$ is antagonistic to the periodic boundary conditions. But here is a set of key observations:
\begin{itemize}
\item The kernel $\langle x_j | Q_\omega|x'_j \rangle$ decays exponentially, on average, as $|x_j-x'_j|$ is increased.
\item When restricting $x_j$ to a finite domain $-N \leq x_j \leq N$ and imposing periodic boundary conditions, we are practically placing the system on the circle $\mathcal C_N$ of perimeter $2N+1$.
\item The factor $x_j -x'_j$ is antagonistic to this circle but we only need to represent it exactly for $x_j$ close to $x'_j$, which leaves plenty of room to make it compatible with the circle (i.e. periodic).
\end{itemize}
Based on these guiding principles, the following procedure was proposed in Ref. [\onlinecite{ProdanAMRX2013bn}]. Let $f:[-1,1] \rightarrow \mathbb R$ be a smooth and periodic function such that $f(r)=r$, for $|r|$ smaller than some $\alpha \lesssim 1$. This function is used to define a function on the circle  $\mathcal C_N$: $f_N(x) = N f(x/N)$, which has the correct periodicity and is equal to $x$ for $|x| <\alpha N$. Let us consider its discrete Fourier representation:
\begin{equation}
f_N(x) = \frac{1}{2N+1}\sum_\lambda c_\lambda \lambda^ x,
\end{equation}
where the sum is over all $2N+1$ solutions of the equation $z^{2N+1}=1$. This will enable us to extend the domain of this function indefinitely (note that $x_j-x'_j$ takes values in the interval $[-2N,2N]$) and to finally define the proper replacement of the antagonistic factor in Eq. (\ref{Comm}):
\begin{equation}
x_j -x'_j \rightarrow \sum_\lambda c_\lambda \lambda^ {x_j -x'_j}.
\end{equation}
From the above approximation, a concrete form of the commutators can be derived. Numerically, we found that the periodicity of the starting function $f$ is not that important in practice, and in most of our calculations we simply use $f(r)=r$ over the entire $[-1,1]$ interval. In this case, the Fourier coefficients are known explicitly and given below.

To summarize, the canonical and optimal finite-volume approximation scheme that emerges from the above arguments consists of substituting the commutator $[X_j, Q_\omega]$ with:
\begin{equation}\label{Commutator}
\boxed{
\lfloor X_j, \widetilde Q_\omega \rfloor = \sum_{\lambda \neq 1} c_\lambda \lambda^{X_j} \widetilde Q_\omega \lambda^{-X_j}, \ \ c_\lambda = \frac{\lambda^{N+1}}{1-\lambda}
}.
\end{equation}
where $\widetilde Q_\omega$ represents the finite-volume approximation of $Q_\omega$, obtained from the restriction $\widetilde H_\omega$ of $H_\omega$ on the finite volume.  Based on the key factors listed above, Ref. [\onlinecite{ProdanAMRX2013bn}] established the following rigorous result. Let $\Phi_j$ be smooth functions and let the accent $\sim$ indicate the restriction to a finite volume, then
\begin{align}
 &\big | \mathcal T\{[ X_{\alpha_1},\Phi_1(H_\omega)] [ X_{\alpha_2},\Phi_2(H_\omega)] \ldots \}  \nonumber\\
 & -\widetilde{\mathcal T}\{\lfloor X_{\alpha_1},\Phi_1(\widetilde H_\omega)\rfloor \lfloor X_{\alpha_2},\Phi_2(\widetilde H_\omega)\rfloor \ldots \} \big | < C_\Phi e^{-\beta N}
\end{align}
where the constant $C_\Phi$ has an explicit expression in terms of the $\Phi_j$ functions. Based on this result, any correlation function involving localized observables and their commutators with the position operators can be canonically approximated on a finite volume, and this approximation converges exponentially fast to the thermodynamic limit. In particular, the canonical finite-volume approximation of the non-commutative winding number in arbitrary odd dimension $D$ is:
\begin{equation}\label{WindingR3}
\nu = \frac{(i \pi)^\frac{D-1}{2}}{D!!} \sum_{\rho \in S_D} (-1)^\rho \ \widetilde{\mathcal T } \left \{\prod_{j=1}^D \widetilde Q_\omega^{-1} \lfloor X_{\rho_j},\widetilde Q_\omega \rfloor  \right \}.
\end{equation}
For convenience, we write the explicit expression in 1-dimension, which is used in the present numerical simulations:
\begin{equation}\label{WindingR4}
\nu = \widetilde{\mathcal T } \left \{ \widetilde Q_\omega^{-1} \lfloor X,\widetilde Q_\omega \rfloor  \right \}.
\end{equation}
In the regime of strong disorder, the quantization of the winding number obtained with this formula is typically exact up to six digits of precision.

\section{Models and Preliminary Analysis}\label{LatticeModels}

We consider the following homogeneous disordered model from the AIII-symmetry class:
\begin{align}\label{Model1}
 (H\bm \psi)_x&= m_x \  \hat \sigma_2  \bm \psi_x \\
& +\nicefrac{1}{2}\ t_x  [(\hat \sigma_1+i \hat \sigma_2 ) \bm \psi_{x+1}+ (\hat \sigma_1-i \hat \sigma_2 ) \bm \psi_{x-1}]\nonumber \\
& +\nicefrac{1}{2} \  t' [( \hat \sigma_1+i \hat \sigma_2 ) \bm \psi_{x+2}+ ( \hat \sigma_1-i \hat \sigma_2 ) \bm \psi_{x-2}],\nonumber
\end{align}
where $\hat \sigma$'s are Pauli's matrices. The disorder is present in the first-neighbor hopping and in the onsite potential:
$$ t_x = t +W_1 \omega_x, \ \ m_x = m+W_2 \omega'_x,$$
where $\omega_x,\, \omega'_x$ are independent randomly generated numbers, uniformly distributed in $\left[-0.5,\,0.5\right]$. The model in Eq. (\ref{Model1})  preserves only the chiral symmetry $\hat \sigma_3  H \hat \sigma_3=-H$, as the first term in Eq. (\ref{Model1}) breaks both the particle-hole ($C=\sigma_3 \mathcal K$) and time-reversal ($T=\mathcal K$, $\mathcal K$= complex conjugation) symmetries. The difference between this model and the one from the previous work\cite{MondragonShemArxiv2013ew} is the second-neighbor hopping term, which enables a richer phase diagram. We have also explored third-neighbor terms and obtained even richer phase diagrams, but those results are not reported here.

\begin{figure}
\includegraphics[width=7.91cm]{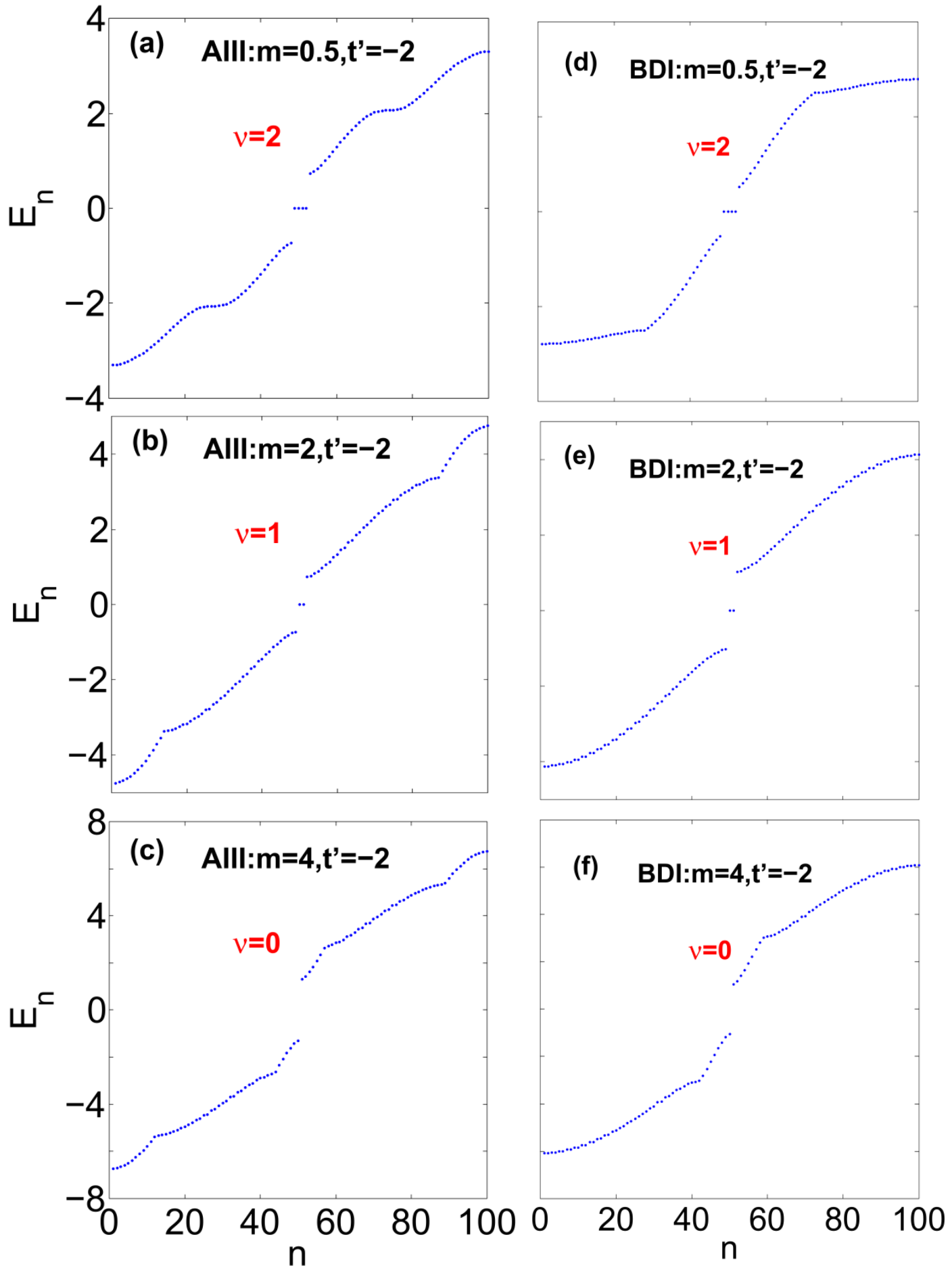}
\caption{(Color online) Plots of the increasingly ordered eigenvalues $E_n$ as a function of $n$ when open boundary conditions are used. This exemplifies the bulk-boundary correspondence principle for the different topological phase from class AIII and class BDI models in a clean limit.}\label{BulkEndstates}
\end{figure}

\begin{figure}
\includegraphics[width=8.6cm]{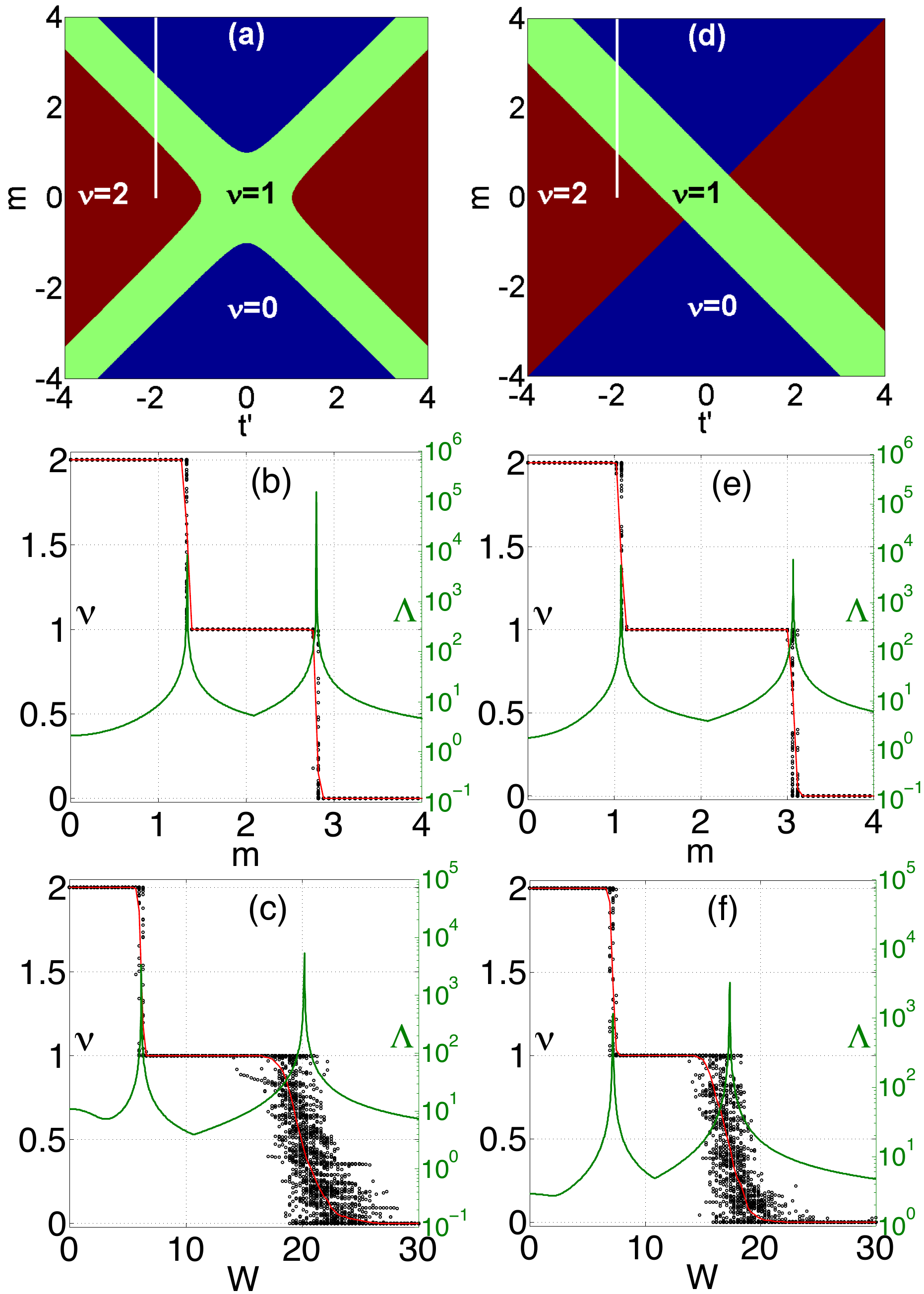}
\caption{(Color online) (a) and (d) The phase diagrams in the phase space $(m,t')$ at $t=1$ and $W_{1,2}=0$. (b) and (e) Maps of the winding number Eq. (\ref{WindingR4}) and of the localization length $\Lambda(E=0)$ along the paths shown in the upper panels (see the light colored segments), with disorder fixed at $2W_1=W_2=2$. (c) and (f) Maps of the winding number Eq. (\ref{WindingR4}) and of the localization length $\Lambda(E=0)$ as a function of disorder $W=2W_1=W_2$, at $t=1$, $m=1$ and $t'=-2$.
The panels in the left column correspond to the AIII symmetric model Eq. (\ref{Model1}), and the ones in the right column to the BDI symmetric model Eq. (\ref{Model2}). The scattered points represent the un-averaged output of Eq. (\ref{WindingR4}) for 100 independent disorder configurations, and the solid (red) line represents the average. The computations for $\nu$ were done with a chain of $N=1000$ unit cells, and in the computation of $\Lambda(E=0)$ the transfer matrix was iterated $10^8$ times.}\label{Nu012}
\end{figure}

\begin{figure}
\includegraphics[width=8.6cm]{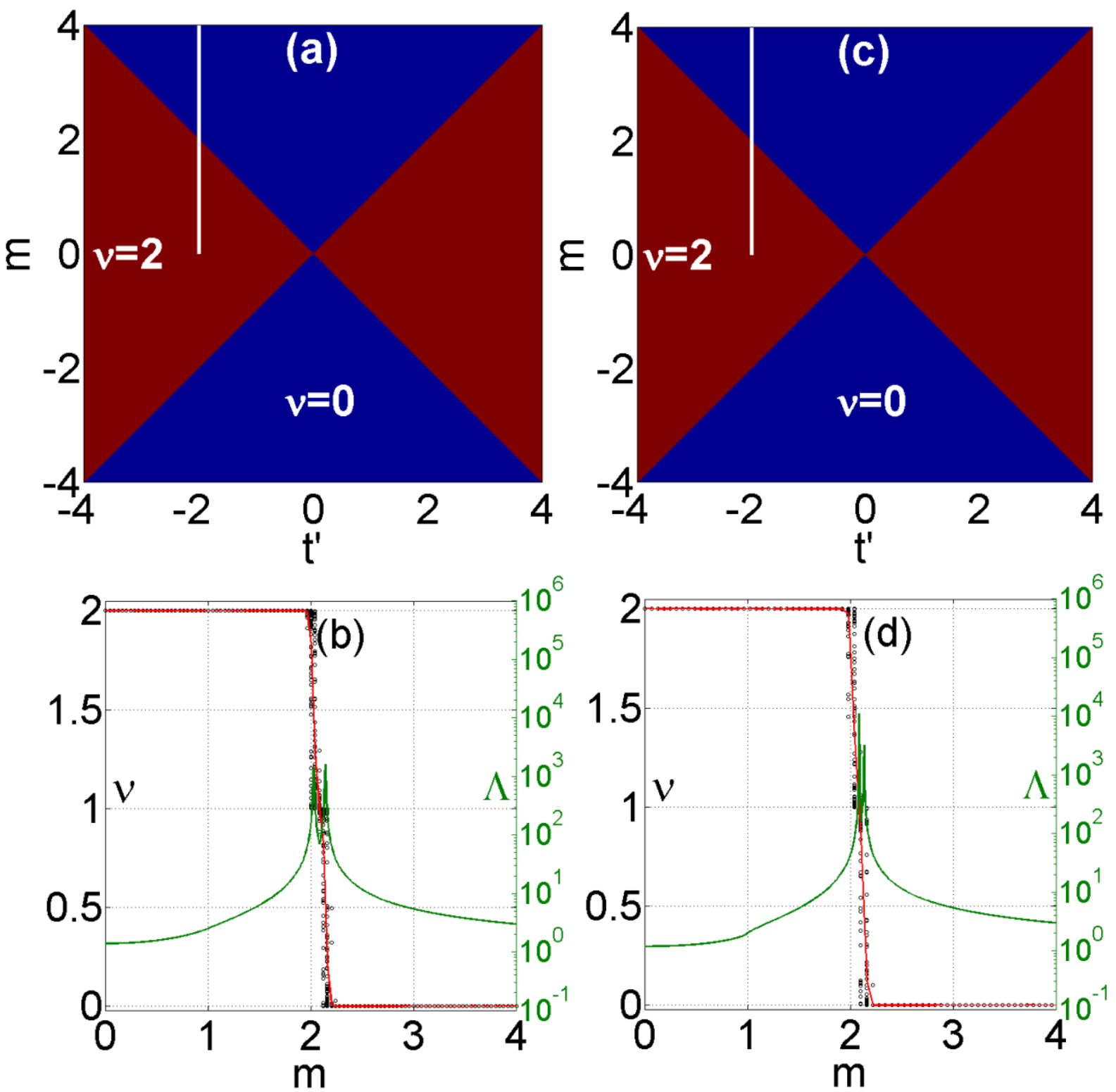}
\caption{(Color online) (a) and (c) The phase diagrams in the phase space $(m,t')$ at $t=0$ and $W_{1,2}=0$. (b) and (d) Maps of the winding number Eq. (\ref{WindingR4}) and of the localization length $\Lambda(E=0)$ along the paths $t'=-2$ shown in the upper panels (see the light colored segments), with disorder fixed at $2W_1=W_2=2$. Everything else is the same as in Fig.~\ref{Nu012}}\label{Nu02}
\end{figure}

It is informative to discuss the phase diagram of the model in the absence of disorder ($W_{1,2}=0$). In this case, the Bloch Hamiltonians take the form:
\begin{equation}
H(k) = \left (
\begin{array}{cc}
0 & t'e^{i2k}+ te^{ik} - i m \\
t'e^{-i2k}+t e^{-ik} +i m & 0
\end{array}
\right ),
\end{equation}
which obeys $\sigma_3H(k)\sigma_3^{-1}=-H(k)$. The winding number of the model can be analytically computed using Eq. (\ref{WindingK1}):
\begin{equation}\label{KWindingNr}
\nu  =\frac{1}{2 \pi i}\oint_{|z|=1} dz \ \frac{2t' z +t}{t' z^2+t z - i m}, \ \ z=e^{ik}.
\end{equation}
The poles of the integrand in the last line are located at:
\begin{equation}
z_{1,2}=\frac{-t \pm \sqrt{t^2 +4imt'}}{2t'},
\end{equation}
and using the residue theorem, one can easily see that:
\begin{equation}
\nu = \left \{
\begin{array}{l}
 0 \ \mbox{if both poles are outside the unit circle}, \medskip \\
1 \ \mbox{if one pole is inside the unit circle}, \medskip \\
 2 \ \mbox{if both poles are inside the unit circle}.
\end{array}
\right .
\end{equation}
The key observation here is that, by fixing some parameters and varying the others, we can generate all phase transitions discussed in Introduction. Indeed, by fixing $t=1$ and varying $t'$ and $m$, we obtain the phase diagram shown in Fig.~\ref{Nu012}(a) where all three phases $\nu=0,1,2$ can be seen. If we set $t=0$, then both poles exit or enter the unit circle synchronously upon varying $m$, leading to the phase diagram shown in Fig.~\ref{Nu02}(a), where only the $\nu=0$ and $\nu=2$ phases can be seen. As is expected, there should be $2$, $1$ and $0$ end states appearing at each end of an open one-dimensional AIII chain for $\nu = 2$, $1$ and $0$, respectively.

By changing the Pauli matrix in the first term of Eq. (\ref{Model1}), from $\hat \sigma_2$ to $\hat \sigma_1$, the symmetry class of the model becomes BDI type:
\begin{align}\label{Model2}
 (H\bm \psi)_x&= m_x \  \hat \sigma_1  \bm \psi_x \\
& +\nicefrac{1}{2}\ t_x  [(\hat \sigma_1+i \hat \sigma_2 ) \bm \psi_{x+1}+ (\hat \sigma_1-i \hat \sigma_2 ) \bm \psi_{x-1}]\nonumber \\
& +\nicefrac{1}{2} \  t' [( \hat \sigma_1+i \hat \sigma_2 ) \bm \psi_{x+2}+ ( \hat \sigma_1-i \hat \sigma_2 ) \bm \psi_{x-2}].\nonumber
\end{align}
Indeed, besides the chiral symmetry, the above model also preserves the time-reversal and charge-conjugation symmetries. The time-reversal operation squares to 1 in this case.

In the absence of disorder ($W_{1,2}=0$), the Bloch Hamiltonians take the form:
\begin{equation}
H(k) = \left (
\begin{array}{cc}
0 & t'e^{i2k}+ te^{ik} + m \\
t'e^{-i2k}+t e^{-ik} + m & 0
\end{array}
\right ),
\end{equation}
which obeys $\sigma_3H(k)\sigma_3^{-1}=-H(k)$. The winding number of the model can be again analytically computed using Eq. (\ref{WindingK1}):
\begin{equation}\label{KWindingNr}
\nu  =\frac{1}{2 \pi i}\oint_{|z|=1} dz \ \frac{2t' z +t}{t' z^2+t z + m}, \ \ z=e^{ik}.
\end{equation}
The poles of the integrand in the last line are located at:
\begin{equation}\label{Poles2}
z_{1,2}=\frac{-t \pm \sqrt{t^2 -4mt'}}{2t'},
\end{equation}
and using the residue theorem again, one can easily see that:
\begin{equation}
\nu = \left \{
\begin{array}{l}
0 \ \mbox{ if both poles are outside the unit circle}, \medskip \\
1 \ \mbox{if one pole is inside the unit circle}, \medskip \\
 2 \ \mbox{if both poles are inside the unit circle}.
\end{array}
\right .
\end{equation}
As such, by fixing some parameters and varying the others, we can generate all phase transitions discussed in Introduction. Indeed, by fixing $t=1$ and varying $t'$ and $m$ we obtain the phase diagram shown in Fig.~\ref{Nu012}(d), where all three phases $\nu=0,1,2$ can be seen. Furthermore, when $m$ is increased, the square root in Eq. (\ref{Poles2}) becomes purely imaginary and both poles exit from the unit circle synchronously upon increasing $m$, enabling us to witness a phase transition between $\nu=0$ and $\nu=2$. Furthermore, by fixing $t=0$, we obtain the phase diagram shown in Fig.~\ref{Nu02}(c) where only the $\nu=0$ and $\nu=2$ phases can be seen.

For a system with chiral symmetry, the bulk-boundary correspondence principle states that at one end of a sample, stable bound states are forming exactly at $E=0$. The number of these stable states is equal to the winding number $\nu$ of the bulk states. In Fig.~\ref{BulkEndstates}, we exemplify this bulk-boundary correspondence principle for all the phases present in the phase-diagrams. Note that the topological phases are practically impossible to be detected by these end states if the insulating gap is filled with a dense localized spectrum.

As one can notice from the phase diagrams at zero disorder, all the AIII and BDI phases can be adiabatically connected. A legitimate question then arises: Are these systems topologically equivalent? To answer, recall that the topological classification is not only about the adiabatic deformation but also about the response of the system against disorder.\cite{SchnyderPRB2008qy,KitaevArXiv2009oh,RyuNJP2010tq} The additional symmetry present in the BDI class leads to considerably different effective $\sigma$-model compared with the AIII class, hence the response of two classes to disorder (which in the BDI case must respect the additional symmetry) can be quite different. One of our goals is to see if this additional symmetry in the BDI model can stabilize bulk extended states.

In the presence of disorder, our tools of analysis are the non-commutative winding number already introduced and discussed above, and the dynamical localization length of the system. The latter is computed using the recursive transfer matrix,\cite{MacKinnonPRL1981vu,MacKinnonZPB1983er} as briefly explained here. Substituting Eqs. (\ref{Model1}) and (\ref{Model2}) into the Schr\"{o}dinger equation $(H\bm \psi)_x=E\bm \psi_x$, we can obtain:
\begin{align}{\label{TransferMatrix}}
&\left (
\begin{array}{c}
\bm\psi_{x+2} \\
\\
\bm\psi_{x+1} \\
\end{array}
\right )
=
\bm T_{x}
\left (
\begin{array}{c}
\bm\psi_{x+1} \\
\\
\bm\psi_{x} \\
\end{array}
\right )\nonumber\\
&=
\left (
\begin{array}{cccc}
-\frac{\eta t_{x+1}}{m_{x+2}}&-\frac{\eta Et_{x}}{ m_{x+2}}&\frac{\eta E^2}{m_{x+2}t'}-\frac{\eta t'}{ m_{x+2}}&-\frac{\eta^2
Em_x}{m_{x+2}t'}\\
\\
0&-\frac{t_x}{t'}&\frac{E}{t'}&-\frac{\eta m_x}{t'}\\
1&0&0&0\\
0&1&0&0\\
\end{array}
\right )
\left (
\begin{array}{c}
\bm\psi_{x+1} \\
\\
\bm\psi_{x} \\
\end{array}
\right ),
\end{align}
where $\eta=1/i$ corresponds to the AIII and BDI symmetry classes, respectively. Due to the existence of the second neighboring hopping $t'$ term, an analytic expression of the localization length at $E = 0$ cannot be obtained as in Ref. [\onlinecite{MondragonShemArxiv2013ew}]. Here, the localization length $\Lambda$ is computed  numerically by:
\begin{equation}{\label{Localization1}}
\Lambda=\frac{1}{\gamma_{\text{min}}},
\end{equation}
where the smallest positive Lyapunov exponent $\gamma_{\text{min}}$ is defined by the eigenvalues $\{e^{\bm\gamma_i};i=1-4\}$ of the matrix,
\begin{equation}
\bm \Gamma=\lim_{N\rightarrow\infty} \left [\prod_{x=1}^{N}\bm T_{x}\prod_{x=N}^{1}\bm T_{x}^\dagger\right ]^{\nicefrac{1}{2N}}.\nonumber
\end{equation}
Note that the numerical method about how to obtain the smallest positive Lyapunov exponent precisely can be found in Refs. [\onlinecite{MacKinnonPRL1981vu}] and [\onlinecite{MacKinnonZPB1983er}]. From Eq. (\ref{Localization1}), if the smallest positive Lyapunov exponent $\gamma_{\text{min}}$ is very close to 0 or $e^{\gamma_{\text{min}}}\sim 1$ there exists a delocalized state, which is indeed observed at $E=0$ at the phase transitions of the AIII and BDI symmetric systems.

\section{Mapping $\nu$ at Strong Disorder}\label{NCWNStrongdisorder}

We first turn on the disorder and keep it fixed at the levels $W_1=1$ and $W_2=2$ while mapping the non-commutative winding number along the paths shown in Figs.~\ref{Nu012}(a) and ~\ref{Nu012}(d) for the AIII and BDI models, respectively. The level of disorder is strong enough to close the insulting gaps of the models throughout these paths.  The results of the numerical calculations are shown in Figs.~\ref{Nu012}(b) and ~\ref{Nu012}(e), which present the output of separate runs for 100 disorder configurations (the scattered data) and the average of $\nu$ over these disorder configurations (the red line). As one can see, the scattered data and the average overlap almost perfectly, which is a direct manifestation of the self-averaging property of the non-commutative formula. Furthermore, $\nu$ can be seen to take the cascade of strict quantized values 2, 1, and 0, for both symmetry classes, with very sharp transitions between the quantized values. The maps of the dynamical localization length along the same paths, shown on the same graphs, reveal Anderson localization-delocalization transitions at each step where $\nu$ changes its quantized values.

Next, we demonstrate that the same cascade of phase transitions can be driven by disorder. For this, we repeat the previous calculations but now we fix the model parameters at $t=1$, $m=1$ and $t'=-2$, which place the system deep in the $\nu=2$ phase, and then we increase the disorder strength in the following fashion $W_2=2W_1=W$, with $W$ running from 0 to the extreme value of 30. The results are shown in Figs.~\ref{Nu012}(c) and \ref{Nu012}(f) for the AIII and BDI symmetric models, respectively, and the cascade of transitions is clearly present. Likewise, here the scattered data show the output of $\nu$ for 100 independent runs with different disorder configurations, and the continuous (red) line represents the disorder average. The self-averaging property is clearly seen at work in these figures, except at the second topological phase transition, but those fluctuations are understandable given the extremely large value of disorder. The numerically calculated localization length can be seen again to diverge at the topological phase transitions.

\begin{figure}
\includegraphics[width=8.6cm]{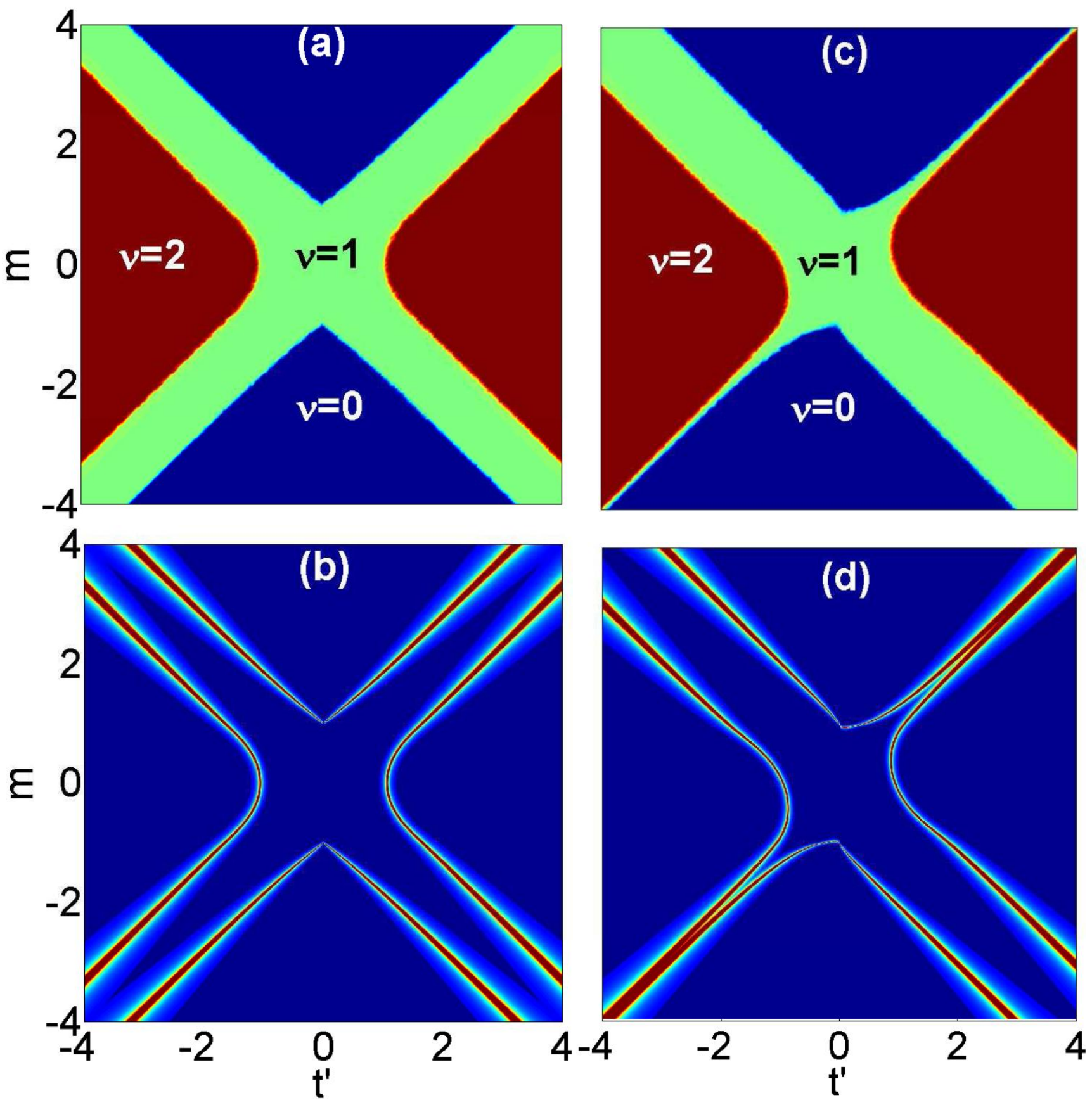}
\caption{(Color online) The maps of the winding number (a) and (c) and localization length (b) and (c) as computed with Eqs. (\ref{WindingR3}) and with the numerical transfer matrix method, respectively, by setting $t=1$ and $W_1=W_2=2$. The computations of $\nu$ were done for $N=1000$  and averaged over 10 disorder configurations. The transfer matrix was iterated $10^8$ times. Note that the figures in the left panel and in the right panel correspond to the class AIII model and the class BDI model, respectively.}\label{PhaseDiagWeakW}
\end{figure}

The calculations so far show that the non-commutative winding number can take more than the values 0 and 1 in the presence of strong disorder. By considering a third-neighbor hopping term we were able to generate topological phases with winding number 3 and this leaves little doubt that $\nu$ can take any integer value. We now investigate the transition from $\nu=2$ directly into $\nu=0$, and we demonstrate that these two phases are distinct. For this, we set the parameter $t$ to zero so that in the clean limit we obtain the phase diagrams shown in Figs.~\ref{Nu02}(a) and ~\ref{Nu02}(c). It is important to remark that in this case there are hoppings only between odd sites, and separately between the even sites, hence we effectively have two disconnected chains. However, we introduce the same disorder as in panels (b) and (c) of Fig.~\ref{Nu012}, $2W_1=W_2=2$, in which case the first-neighbor hoppings are restored and the chains become coupled through disorder. We map $\nu$ along the paths $\Gamma'_1$ and $\Gamma'_2$, and the results for the two models are shown in Figs.~\ref{Nu02}(b) and ~\ref{Nu02}(d), together with a map of the dynamical localization length. The Anderson localization-delocalization is clearly visible, which is a strong indication that the $\nu=2$ and $\nu=0$ phases are indeed distinct.

\begin{figure}
\includegraphics[width=8.6cm]{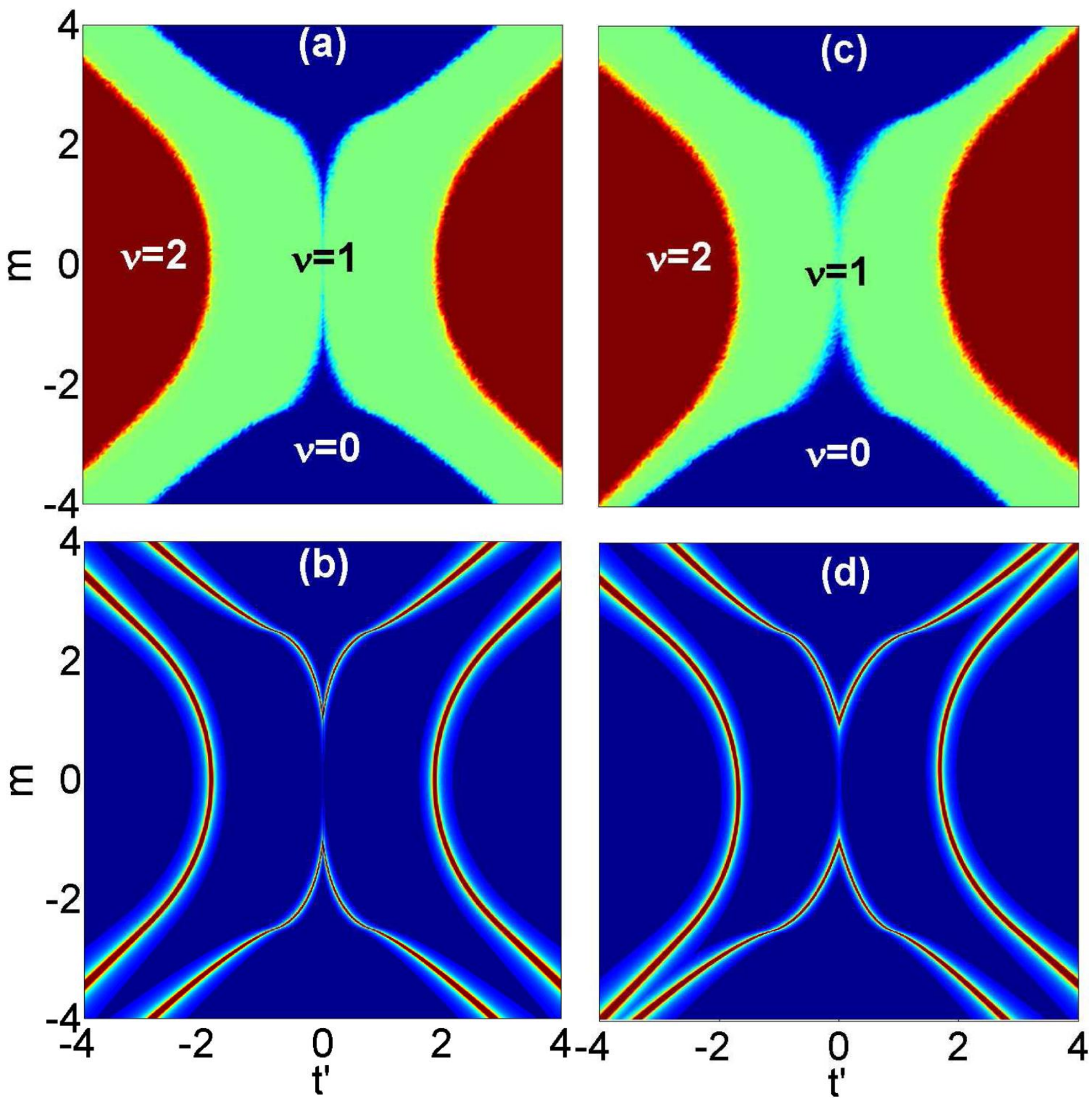}
\caption{(Color online) Same as Fig.~\ref{PhaseDiagWeakW}, except for $W_1=W_2=5$.}\label{PhaseDiagStrongW}
\end{figure}

Lastly, we have combed the energy spectrum and found that the dynamical localization length is always finite, except at the Anderson localization-delocalization phase transitions discussed above. This shows that for both AIII and BDI systems, the quantized topological invariant is carried entirely by localized states.

\section{Analysis of the Phase Diagrams}\label{PhaseDiagram}

Our analysis continues with computations of the phase diagrams, which will reveal that the phases characterized by different $\nu$'s are entirely disconnected from each other and that they are separated by an Anderson localization-delocalization transition. Furthermore, the phase diagrams for both models display an intricate behavior as a functions of disorder. The phase boundaries move non-monotonically with the disorder, with the domain of the non-trivial topological phases increasing at the beginning  and then decreasing until the domains disappear. As such, this provides another example in which the disorder alone can drive the system from a trivial to a topological phase. \cite{LiPRL2009xi,HJiangPRB2009pd,GrothPRL2009xi,HMGuoPRL2010pj,ProdanPRB2011vy,JTSongPRB2012pd,YYZhangPRB2012gw}

The phase diagrams of the models in the $(t',m)$ plane and at a finite disorder $W_1=W_2=2$ are shown in Fig.~\ref{PhaseDiagWeakW}. Specifically, the maps of the non-commutative winding number are reported in panel (a) for the AIII model and in panel (c) for the BDI model. Comparing the first map with the one obtained at zero disorder, one can immediately observe a relative stability of the phases, but when comparing the second map an instability can be witnessed in panel (c), caused by the spill of phase $\nu=1$ in between the phases $\nu =0$ and $\nu=2$. This is interesting because it shows that the disorder alone can transform the $2 \rightarrow 0$ transition (which is very interesting by itself)  into the monotonic sequence of transitions $2 \rightarrow 1 \rightarrow  0$. Panels (b) and (d) report the maps of the localization length $\Lambda(E=0)$, showing fine lines where $\Lambda(E=0)$ diverges and which coincide perfectly (within the numerical accuracy) with the boundaries between different $\nu$-phases. This confirms that the topological phases (identified by the values of $\nu$) are separated by an Anderson localization-delocalization transition throughout the phase boundaries.  We have repeated the calculations with a stronger disorder $W_1=W_2=5$ and the results are reported in Fig.~\ref{PhaseDiagStrongW}. Here one can see that the phase diagrams have been distorted quite pronouncedly by the disorder, and the protrusion of the $\nu=1$ phase in between the $\nu=1,2$ phases has been accentuated. In fact, the phase diagrams for the two models look very alike, which is expected because in the regime of infinite disorder the two models are similar. The maps of the dynamical localization length at $E=0$ show again that all the topological phases are separated from each other by an Anderson localization-delocalization transition.

\begin{figure}
\includegraphics[width=8.2cm]{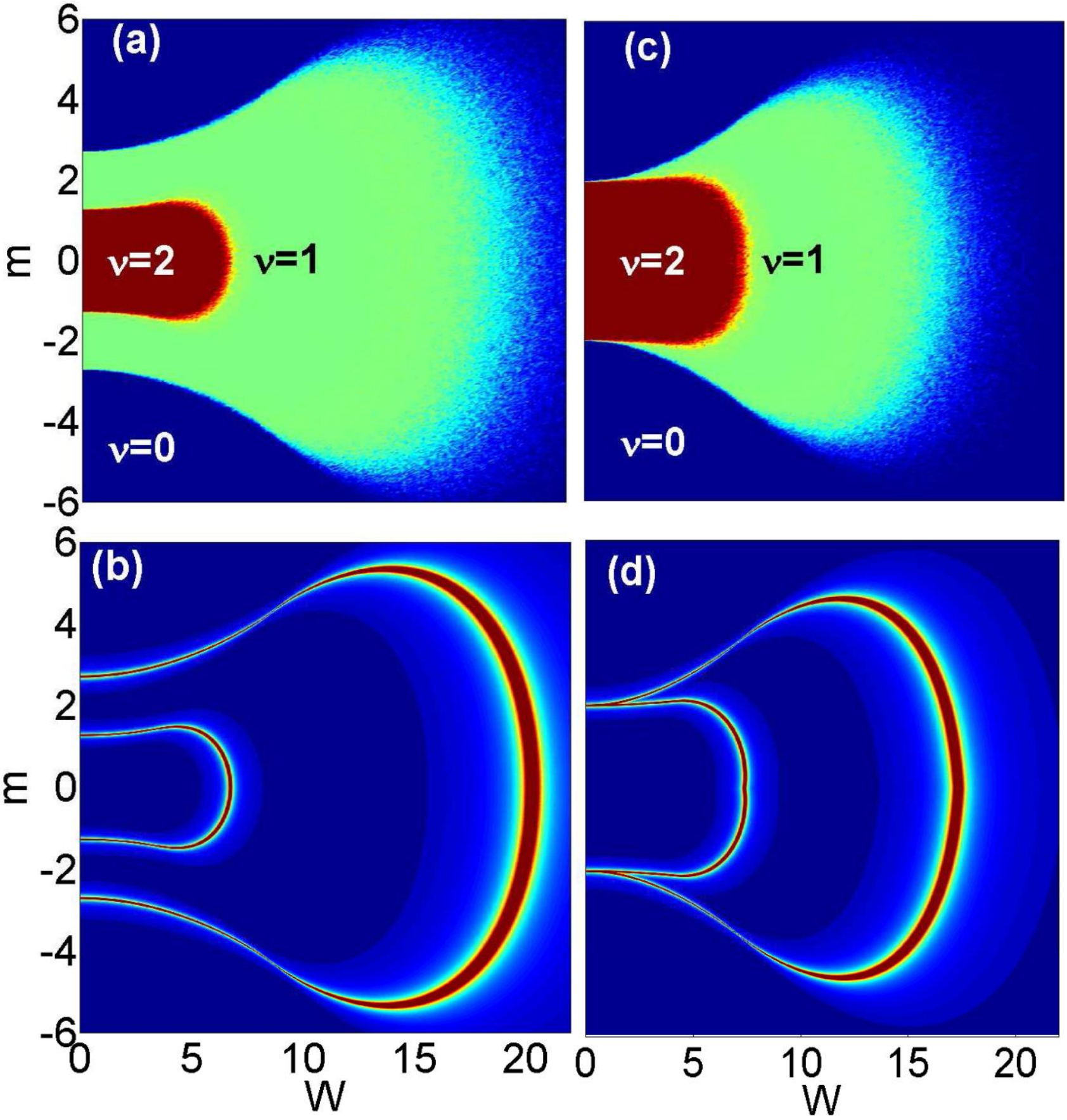}
\caption{(Color online) The maps of the winding number (a) and (c) and localization length (b) and (d) as computed with Eqs. (\ref{WindingR3}) and with the numerical transfer matrix method, respectively, by setting $t=1$, $t'=2$ and $W=2W_1=W_2$. The computations of $\nu$ were done for $N=1000$  and averaged over 10 disorder configurations. The transfer matrix was iterated $10^8$ times. Note that the figures in the left panel and in the right panel correspond to the class AIII model and the class BDI model, respectively.}\label{PhaseDiag1VsW}
\end{figure}

\begin{figure}
\includegraphics[width=8.5cm]{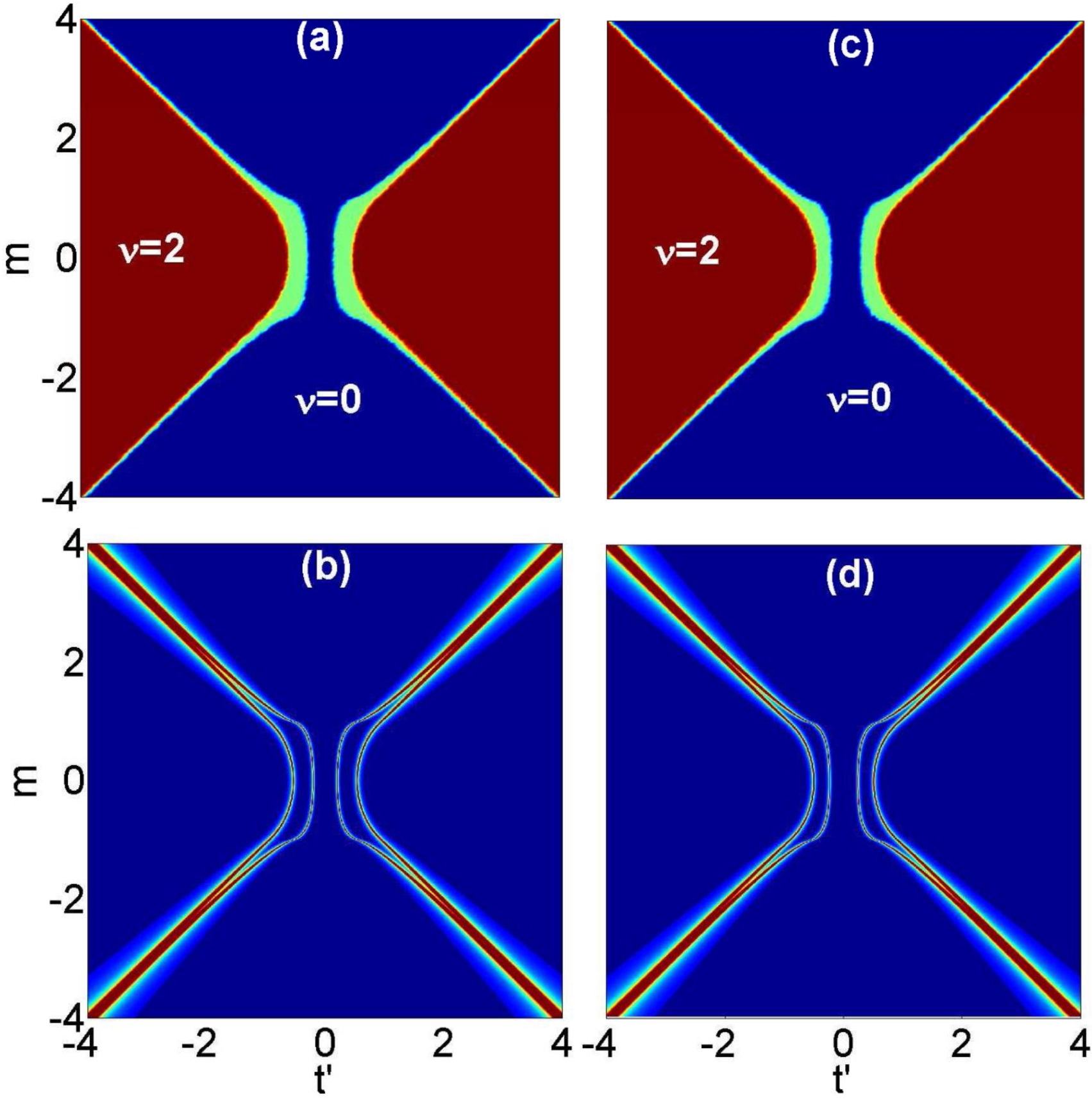}
\caption{(Color online) The maps of the winding number (a) and (c) and localization length (b) and (d) as computed with Eqs. (\ref{WindingR3}) and with the numerical transfer matrix method, respectively, by setting $t=1$, $t'=2$ and $2W_1=W_2=2$. The computations of $\nu$ were done for $N=1000$  and averaged over 10 disorder configurations. The transfer matrix was iterated $10^8$ times. Note that the figures in the left panel and in the right panel correspond to the class AIII model and the class BDI model, respectively.}\label{PhaseDiag2VsW}
\end{figure}

The phase diagrams in the $(W,m)$ plane, with $W=2W_1=W_2$, are shown in Fig.~\ref{PhaseDiag1VsW}. Here one can explicitly see the topological phases growing as the disorder is turned on, and as a result the topological domain with $\nu=1$ spills into the domain formerly occupied by the trivial phase, and the domain with $\nu=2$ spills into the domain formerly occupied by the $\nu=1$ phase. For example, hypothetical materials described by the two models with $m=3$, would be trivial in the clean limit but become topological insulators when increasing the disorder, as shown in Figs. \ref{PhaseDiag1VsW}(a) and \ref{PhaseDiag1VsW}(c). Furthermore, for the BDI model, the phase $\nu=1$ is entirely absent in the clean limit, but it robustly emerges after the disorder is turned on and fully develops at extreme values of disorder. The maps of the dynamical localization length continues to show that all topological phases are separated by an Anderson localization-delocalization transition.

\begin{figure}
\includegraphics[width=8cm]{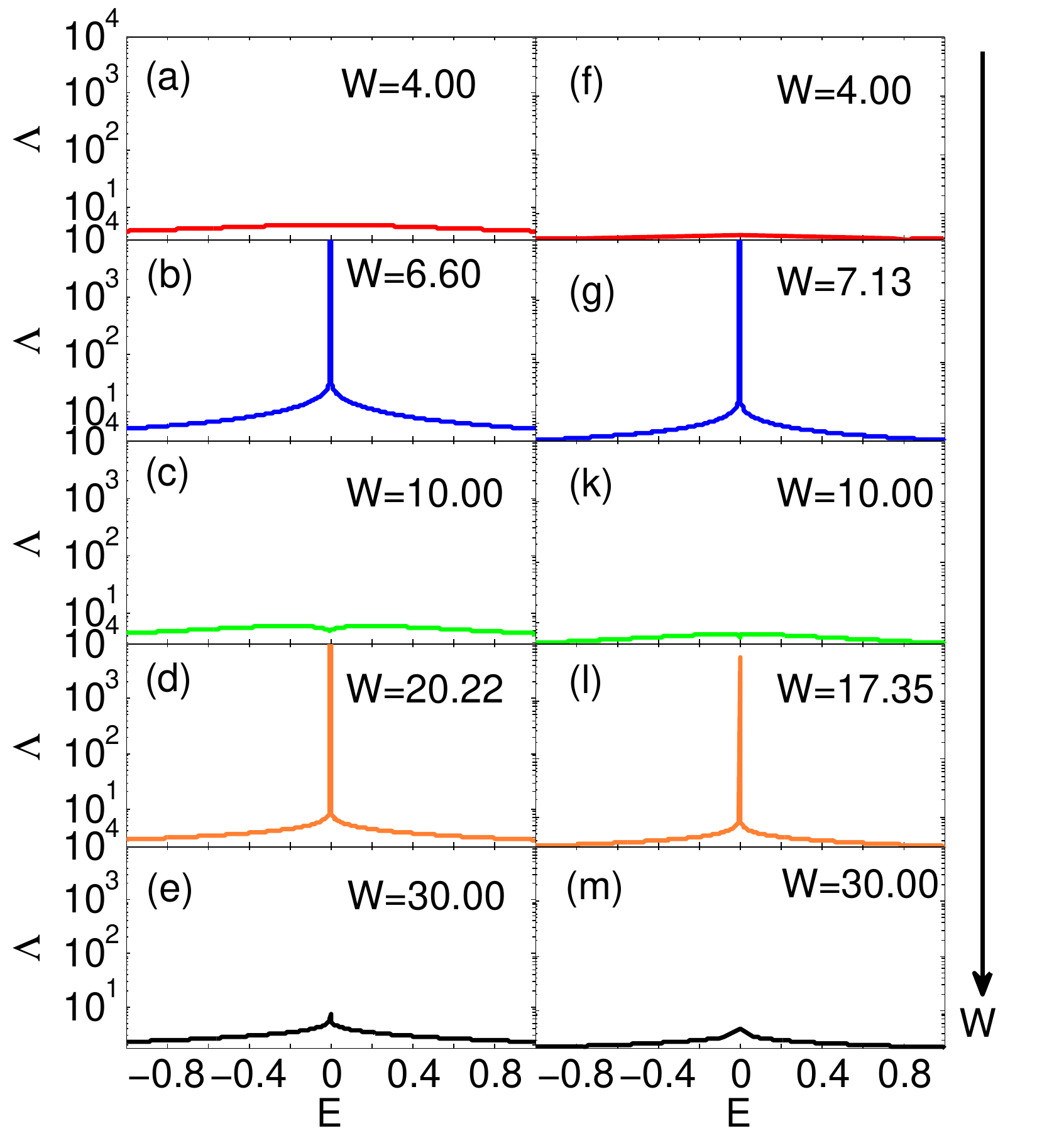}
\caption{(Color online) Maps of the localization length as a function of energy, for five points of the phase diagram reported in Figs.~\ref{PhaseDiag1VsW}(a) and ~\ref{PhaseDiag1VsW}(c), all taken at $t=1$, $t'=-2$ and $m=0.5$. The level of disorder is indicated on each panel. The left and right panels correspond to the class AIII model and the class BDI model, respectively.}\label{LocLength}
\end{figure}

Next, we have considered the phase diagrams from Fig.~\ref{Nu02}, which are for the clean limit and show only the phases $\nu=2$ and $\nu=0$, and we introduced a disorder $2W_1=W_2=2$. The phase diagram in the plane $(m,t')$ and in the presence of this disorder is reported in Fig.~\ref{PhaseDiag2VsW}. Here we see again the phase $\nu=1$ emerging again between the $\nu=0$ and $\nu=2$ phases, even if it was not there to start with. It becomes clearer now that the transition $\nu =2 \rightarrow \nu = 0$ is not stable and instead it seems that this transition is always broken to the cascade of transitions $\nu =2 \rightarrow \nu=1 \rightarrow \nu = 0$ when the disorder is introduced. Nevertheless, if $\nu=0$ and $\nu=2$ are identical phases, then inherently we would have encountered many instances where a crossing from $\nu=2$ into $\nu=0$ happens without an Anderson localization-delocalization transition. However, from the maps of the dynamical localization length reported in the lower panels of Fig.~\ref{PhaseDiag2VsW}, one can see that an Anderson transition is always encountered when crossing from $\nu=0$ phase to $\nu =2$ phase and vice versa. This leaves very little doubt that these two phases are indeed distinct.

Lastly, in Fig.~\ref{LocLength} we report traces of the dynamical localization length as a function of energy, taken at five points of the phase diagrams reported in Fig.~\ref{PhaseDiag1VsW}. The first, third, and fifth points are located inside the $\nu=2$, $\nu=1$ and $\nu=0$ phases, respectively, while the second and fourth points are located on the phase boundary between the $\nu =2$ and $\nu =1$ phases, and between the $\nu=1$ and $\nu=0$ phases, respectively. The computations were performed for the AIII (left panel) as well as the BDI (right panel) models. The important observation here is that, for all points that are away from the phase boundary, the localization length is finite, proving the absence of the extended states in the spectrum, even for the topological phases. This is a direct confirmation that the topological winding number is carried entirely by localized states. Furthermore, one can see that the divergence of the localization length at the boundary between the phases develops strictly at $E=0$, and no trace of the so called levitation and annihilation phenomenon can be detected.

\section{Conclusions}\label{Conclusions}

The non-commutative winding numbers for one-dimensional disordered phases from the AIII and BDI symmetry classes have been explicitly shown to take several integer values, and that a phase characterized by a specific value of the winding number is enclosed in a phase-boundary where an Anderson localization-delocalization transition takes place.  In particular, the phases characterized by $\nu=0$ and $\nu=2$ have been shown to be distinct, a fact that rules out a $\mathbb Z_2$-classification and supports a $\mathbb Z$-classification for these systems. This also shows that the electric polarization, which takes the quantized values $0$ and $\frac{1}{2}$, does not provide a full classification for these phases.

We have also demonstrated explicitly that the energy spectra of the topological phases are fully localized, the topological invariants are carried entirely by localized states, and as a consequence, the Anderson localization-delocalization transition develops entirely from this localized spectrum, without the well-known mechanism of pair levitation and annihilation.

The numerical algorithms for the non-commutative winding number and their performance were discussed in detail. In all our simulated phase diagrams, the quantization of the topological invariant in the presence of extreme disorder is extremely accurate to more than six digits of precision. The self-averaging property, which was also explicitly demonstrated, makes the non-commutative winding number one of the most effective tools for the analysis of topological systems with chiral symmetry.

\acknowledgments We would like to thank Ian Mondragon-Shem and Taylor L. Hughes for insightful discussions. This work was supported by the U.S. NSF grants DMS-1066045, DMR-1056168, NSFC under grants No. 11204065 and NSF-Hebei Province under grants No. A2013205168.

\end{document}